\documentclass[amsmath,amssymb,prc,superscriptaddress,epsfig,aps,twocolumn,eqsecnum,showpacs]{revtex4}
\usepackage{graphicx}  %for graphics
\usepackage{dcolumn}% Align table columns on decimal point
\usepackage{bm}% bold math

\newcommand{\mb}[1]{\mathbf{#1}}
\newcommand{\eq}[1]{\begin{equation} #1 \end{equation}}

\newcommand{\eqa}[1]{\begin{eqnarray} #1 \end{eqnarray}}

\begin{document}

\title{Pairing renormalization and regularization
within the local density approximation}

\author{P.J.~Borycki}
\affiliation{Department of Physics \& Astronomy, University of
Tennessee, Knoxville, Tennessee 37996, USA}  \affiliation{Institute
of Physics, Warsaw University of Technology, ul. Koszykowa 75,
00-662~Warsaw, Poland}

\author{J.~Dobaczewski}
\affiliation{Department of Physics \& Astronomy, University of
Tennessee, Knoxville, Tennessee 37996, USA} \affiliation{Physics
Division,  Oak Ridge National Laboratory, P.O. Box 2008, Oak Ridge,
Tennessee 37831, USA}  \affiliation{Institute of
Theoretical Physics, Warsaw University, ul. Ho\.za 69, 00-681
Warsaw, Poland}\affiliation{Joint Institute for Heavy-Ion
Research, Oak Ridge, Tennessee 37831, USA}

\author{W.~Nazarewicz}
\affiliation{Department of Physics \& Astronomy, University of
Tennessee, Knoxville, Tennessee 37996, USA} \affiliation{Physics
Division,  Oak Ridge National Laboratory, P.O. Box 2008, Oak Ridge,
Tennessee 37831, USA}  \affiliation{Institute of
Theoretical Physics, Warsaw University, ul. Ho\.za 69, 00-681
Warsaw, Poland}

\author{M.V.~Stoitsov}
\affiliation{Department of Physics \& Astronomy, University of
Tennessee, Knoxville, Tennessee 37996, USA} \affiliation{Physics
Division,  Oak Ridge National Laboratory, P.O. Box 2008, Oak Ridge,
Tennessee 37831, USA} \affiliation{Joint Institute for Heavy-Ion
Research, Oak Ridge, Tennessee 37831, USA} \affiliation{Institute of
Nuclear Research  and Nuclear Energy, Bulgarian Academy of Sciences,
Sofia-1784, Bulgaria}

\date{\today}

\begin{abstract}
We discuss methods used in mean-field theories to treat pairing
correlations within the local density approximation. Pairing renormalization and regularization procedures are compared in spherical and deformed nuclei. Both
prescriptions give fairly similar results, although the theoretical
motivation, simplicity, and stability of the regularization procedure
makes it a method of choice for future applications.
\end{abstract}


\maketitle

\section{Introduction}\label{sec_intro}

One of the main goals of the low energy nuclear theory is to build a comprehensive
microscopic framework, in which nuclear bulk properties,
excitations, and low energy reactions can be described. For medium-mass and heavy nuclei, self-consistent methods based on the Density Functional
Theory (DFT) \cite{[Hoh64],[Koh65]} have already achieved a level of precision that allows
for analysis of experimental data for a wide range of properties of
nuclei throughout the chart of the nuclides. For example, the self-consistent
Hartree-Fock-Bogoliubov (HFB)
models based on the Skyrme energy functionals \cite{[Sky56],[Vau72],[Per04]}
are able nowadays to reproduce
nuclear masses with an rms error of about 700\,keV \cite{[Sam02],[Gor02]}. The
development of a universal nuclear density functional, however,
still requires a better understanding and improved description of the density dependence,
isospin effects, pairing force, many-body correlations, and symmetry restoration.

Nuclear pairing is an important ingredient of the nuclear density functional, and it becomes crucial for open shell nuclei, in particular weakly
bound systems, where the effects of coupling to continuum become significant
\cite{[Dob96],[Dob01]}. In
this case, the BCS model is not adequate \cite{[Dob96]} and the fully
self-consistent HFB approach must be used.

In most HFB applications, pairing interaction is assumed to be either in the form of the finite range Gogny force \cite{[Dec80]} or the
zero-range, possibly density-dependent, delta force \cite{[Cha76],[Dob96],[Dob01]}.
Gogny interaction in the pairing channel can be viewed as a regularized contact interaction, with regularization fixed through the finite range.
The resulting pairing field is, however, nonlocal.

Calculations using the contact interaction are numerically simpler, but one has to apply a cutoff procedure within a given
space of single-particle (s.p.) states \cite{[Dob84],[Dob96]}. When the dimension of this space increases, the
pairing gap diverges for any given strength of the interaction.
Therefore, the pairing strength has to be readjusted for each
s.p.\ space. Thus the energy cutoff and the pairing strength together define
the pairing interaction, and this definition can be understood as a phenomenological introduction of finite range \cite{[Dob96],[Esb97]}. Such a procedure is usually referred to as the renormalization of the
contact pairing force. It is performed in the spirit of the effective
field theory, whereupon contact interactions are used to describe
low energy phenomena while the coupling constants are readjusted for
any given energy cutoff to take into account neglected high energy
effects.

%The use of a contact interaction in a limited space determined by a cutoff in the s.p.\ space can be understood as a phenomenological introduction of finite range \cite{[Dob96],[Esb97]}. The effective range of the interaction is determined by the energy cutoff while the strength parameter is chosen to reproduce accordingly the empirical pairing gaps or scattering length.

The renormalization procedure for the zero-range pairing interactions has
been explored in Ref. \cite{[Dob96]} using the numerical
solutions of the HFB equations. It has been shown that by renormalizing
the pairing strength for each value of the cutoff energy one
practically eliminates the dependence of the HFB energy on the
cutoff parameter.

Recently, the issue of contact pairing force has been addressed in
Refs.\ \cite{[Mar98],[Pap99],[Bru99],[Bul99],[Bul01],[Bul02],[Bul02a],[Yu-03],[Bul04]}, suggesting that the
renormalization procedure can be replaced by a
regularization scheme which removes the cutoff energy
dependence of the pairing strength. In subsequent papers, this regularization
scheme has been applied to properties of the infinite nuclear matter
\cite{[Bul02]}, spherical nuclei
\cite{[Yu-03],[Nik05]}, and trapped fermionic atoms \cite{[Bru99],[Gra03]}.

In this study, we investigate the stability of the regularization
scheme with respect to the
cutoff energy for both spherical and deformed
nuclei. Differences between the HFB results emerging from the pairing
renormalization and pairing regularization procedures are analyzed.

The HFB and Skyrme HFB formalisms
have been explained in great detail in many papers (see, e.g., Refs.
\cite{[Rin80],[Dob84]}). The notation used in the present paper is consistent
with that of Refs.\ \cite{[Dob84],[Dob96],[Sto05]}. This work is organized as follows.
Sec.~\ref{ssec} gives a
brief introduction to the pairing renormalization and regularization
schemes. In Sect.~\ref{sec_results} we explain the numerical framework used.
The comparison between pairing regularization and
renormalization techniques, studied for a large set of
spherical and deformed nuclei, is discussed in Sec.~\ref{chains}.
Finally, the summary and conclusions are given in Sec.~\ref{sec_summary}.

\section{The cutoff procedures}\label{ssec}

%Within the Skyrme HFB approach, the low-energy nuclear structure phenomena are described by using nucleonic degrees of freedom. The relevant energy scale is, therefore, related to typical binding energies of s.p.\ states, which extend between about 50\,MeV for most strongly bound nucleons down to about 10\,MeV in stable nuclei, or to zero in nuclei at the drip lines. The energy scale related to pairing phenomena is significantly smaller, of the order of several pairing gaps, i.e., about 5\,MeV around the Fermi surface. However, within the self-consistent approach, such as HFB, the s.p.\ and pairing fields are strongly coupled; therefore, the larger energy scale of about 50\,MeV should be used when defining the energy cutoff.

\subsection{Pairing Renormalization Procedure}\label{sec_renorm}

Within the HFB theory, the energy cutoff
can be applied either to the s.p.\ or to the quasiparticle
spectrum. The first option is used when the HFB equations
are solved within a restricted s.p.\ space. However, the s.p.\ energies
play only an auxiliary role in the HFB method, and the cutoff
applied to the quasiparticle spectrum is more justified. This is done by
using the so-called equivalent s.p.\ spectrum
\cite{[Dob84]}:
\eq{\bar{e}_{n}=(1-2P_{n})E_{n}+\mu,}
where $E_{n}$ is the quasiparticle energy and $P_{n}$ denotes
the norm of the lower component of the HFB wave function.

 Due to
the similarity between $\bar{e}_{n}$ and the s.p.\
energies, one takes into account only those quasiparticle states
for which $\bar{e}_{n}$ is less than the assumed cutoff energy
$\epsilon_{\text{cut}}$.

It was shown \cite{[Dob96]} that for
a fixed pairing strength the pairing
energies depend significantly on the energy cutoff.
Within the renormalization scheme employed in this work, we use the prescription of
adjusting the pairing strength to obtain a fixed average neutron pairing gap \cite{[Dob84]},
\eq{\bar{\Delta}=-\frac{1}{N}\int d^3\mb{r}d^3\mb{r'}\sum_{\sigma\sigma'}\tilde h(\mb{r}\sigma,\mb{r'}\sigma')\rho(\mb{r'}\sigma',\mb{r}\sigma),\label{eq:deltabar}}
in $^{120}$Sn equal to the experimental value of
1.245\,MeV. In Eq.~(\ref{eq:deltabar}) $N$ is the number of particles, $\rho$ is the particle density, and $\tilde h$ is the pairing Hamiltonian (see Appendix~\ref{app:divergence}).

Such a procedure almost eliminates the dependence of the HFB energy on the cutoff \cite{[Dob96]}.

\subsection{Pairing Regularization Procedure}\label{sec_method}

Using the HFB equations and properties of the Bogoliubov transformation
(see appendix \ref{app:divergence} for details), one concludes that
the local abnormal density $\tilde\rho$ has a singular behavior when
$\epsilon_{\text{cut}}\rightarrow\infty$. The standard regularization
technique is to remove the divergent part and define the regularized
local abnormal density
$\tilde\rho_r(\mb{r})$ as
\eq{\tilde\rho_r(\mb{r})=\lim_{\mb{x}\rightarrow 0}
\left[\tilde\rho(\mb{r}-\mb{x}/2,\mb{r}+\mb{x}/2)-f(\mb{r},\mb{x})\right],}
where $f$ is a regulator which removes the divergence at $\mb{x}=0$.

For cutoff energies high enough, one can explicitly identify
\cite{[Bul99],[Bul01],[Bul02]} components generating divergence in
the abnormal density (see, e.g., Eq.~(21) of Ref. \cite{[Bul01]}):
\eq{f(\mb{r},\mb{x})=\frac{i\tilde h(\mb{r})M^*(\mb{r})k_F(\mb{r})}{4\pi\hbar^2}
+\frac{\tilde h(\mb{r})}{2}G_{\mu}(\mb{r}+\mb{x}/2,\mb{r}-\mb{x}/2),\label{eq:frx}}
where $G_{\mu}$ is the
s.p.\ Green's function at the Fermi level $\mu$ in the
truncated space, $M^*$ is the effective mass, and the Fermi momentum
is
\eq{k_F(\mb{r})=\frac{\sqrt{2M^*(\mb{r})}}{\hbar}\sqrt{\mu-U(\mb{r})},\label{eq:k_f}}
with  $U$ being the self-consistent mean-field potential.

The first term in Eq.\ (\ref{eq:frx})
comes from the MacLaurin expansion with respect to $\mb{x}$; it
guarantees that the regularization procedure does not introduce any
constant term to the abnormal density and that
$f(\mb{r},\mb{x})$ solely represents the divergent part of
$\tilde\rho$.

Using the Thomas-Fermi approximation, the local s.p.\
Green's function $G_{\mu}(\mb{r}):=G_{\mu}(\mb{r},\mb{r})$ becomes
\cite{[Bul01],[Bul02]}
\eq{
G_{\mu}(\mb{r})=\frac{1}{2\pi^2}\lim_{\gamma\rightarrow
0}\int_0^{k_\text{cut}(\mb{r})}\frac{k^2dk}{\mu-\frac{\hbar^2k^2}{2M^*(\mb{r})}-U(\mb{r})+i\gamma},\label{eq:gf}}
where the cutoff momentum is given by:
\eq{k_{\text{cut}}(\mb{r})=\frac{\sqrt{2M^*(\mb{r})}}{\hbar}
\sqrt{\epsilon_{\text{cut}}+\mu-U(\mb{r})}.}

The regularized pairing Hamiltonian and the pairing energy density
may be written, respectively, as \cite{[Bul01]}:
\eqa{
\tilde h(\mb{r})&=&g(\mb{r})\tilde\rho_r(\mb{r})=g_{eff}(\mb{r})\tilde\rho(\mb{r})\\
{\mathcal H}_{pair}(\mb{r})&=&\frac{1}{2}g_{eff}(\mb{r})\tilde\rho(\mb{r})^2,\label{eq:EDF_pairing}}
where the effective pairing strength \cite{[Bul99],[Bul01],[Bul02]},
\eq{g_{eff}(\mb{r})=\left(\frac{1}{g(\mb{r})}+\frac{G_{\mu}(\mb{r})}{2}
+\frac{iM^*(\mb{r})k_F(\mb{r})}{4\pi\hbar^2}\right)^{-1},}
after calculating integral (\ref{eq:gf}), can be expressed in the form:
\begin{widetext}
\eq{
g_{eff}(\mb{r})=\left\{
\begin{array}{lc}
\left[\frac{1}{g(\mb{r})}-\frac{M^*(\mb{r})
k_\text{cut}(\mb{r})}{2\pi^2\hbar^2}\left(1-\frac{k_F(\mb{r})}{2 k_\text{cut}(\mb{r})}\ln
\frac{k_\text{cut}(\mb{r})+k_F(\mb{r})}{k_\text{cut}(\mb{r})-k_F(\mb{r})}\right)\right]^{-1}&k_F(\mb{r})^2\ge 0
\\~\\
\left[\frac{1}{g(\mb{r})}-\frac{M^*(\mb{r})
k_\text{cut}(\mb{r})}{2\pi^2\hbar^2}\left(1+\frac{|k_F(\mb{r})|}{k_\text{cut}(\mb{r})}
\arctan\frac{|k_F(\mb{r})|}{k_\text{cut}(\mb{r})}\right)\right]^{-1}&k_F(\mb{r})^2<0
\end{array}\right..\label{eq:g^ri}
}
\end{widetext}

In this regularization scheme, only the Green's function is calculated
using the Thomas-Fermi approximation. The densities, potentials, and
chemical potential are determined self-consistently within the HFB
theory. Consequently, the Fermi momentum (\ref{eq:k_f}) depends on
microscopic HFB quantities. According to the sign of $k_F^2$, one
of the expressions (\ref{eq:g^ri}) is used.

In Ref.~\cite{[Bul02a]} a different regularization scheme has been proposed that
involves truncation below and above the Fermi level. However, the HFB
calculations in the quasiparticle basis should be performed for a
high cutoff energy of 50\,MeV and higher \cite{[Dob96]}. Since the
magnitude of the self-consistent mean field $U$ is also about
50\,MeV, for such a high cutoff energy both methods are equivalent.
The Thomas-Fermi approximation
requires that, in order to obtain results independent of
$\epsilon_{\text{cut}}$, its value should be high enough for
$k_\text{cut}$ to be real everywhere.

Through the density dependence of $g_{eff}$,
$k_\text{cut}$, and $k_F$, there appear
rearrangement terms in the self-consistent mean-field potential:
\eqa{\lefteqn{\frac{\delta \mathcal H_{pair}}{\delta\rho}=
\frac{\delta g_{eff}}{\delta \rho}\tilde\rho^2=\tilde\rho^2\times}&&\nonumber\\
&&\times\left(\frac{\partial g_{eff}}{\partial g}\frac{\delta g}{\delta \rho}+
\frac{\partial g_{eff}}{\partial k_F}\frac{\delta k_F}{\delta \rho}+
\frac{\partial g_{eff}}{\partial k_\text{cut}}
\frac{\delta k_\text{cut}}{\delta \rho}\right).\hspace{8pt}\label{eq:rearr} }
The first term in Eq.~(\ref{eq:rearr}) is similar to the usual
rearrangement term, while the other two terms associated with the
regularization procedure are entirely new. It is easy to check that
all the terms appearing in Eq.~(\ref{eq:rearr}) are continuous at the
classical turning point $k_F(\mb{r})=0$.

\begin{figure}
\begin{center}
\includegraphics[width=0.95\columnwidth]{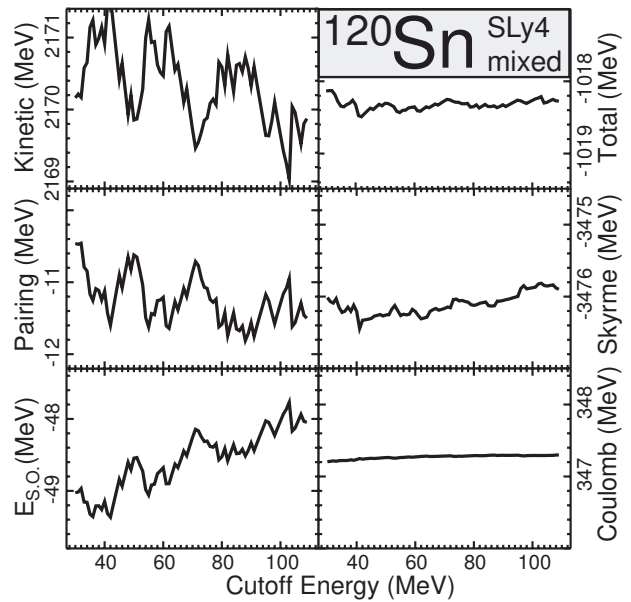}
\end{center}
\caption{Various contributions to the HFB energy for $^{120}$Sn as a function of $\epsilon_{\text{cut}}$. Calculations are
performed using the SLy4 Skyrme functional and mixed pairing interaction (\ref{edhppd}).
}
\label{fig:energy_contribs_fluct.eps}
\end{figure}

In Eq.~(\ref{eq:EDF_pairing}), the pairing energy density is divergent
with respect to the cutoff energy. However, the pairing energy
itself is not an observable, and in order for the energy density
functional to be independent of the cutoff, other terms have to cancel out this divergence. As discussed in Refs. \cite{[Mar98],[Pap99],[Bul99],[Bul01]}, the kinetic energy density $\tau$ has the same type of
divergence as the abnormal density $\tilde\rho$, and the sum
\eq{\mathcal{H}_{kin+pair}(\mb{r})=-\frac{\hbar^2}{2M^*(\mb{r})}\tau
+\frac{1}{2}g_{eff}(\mb{r})\tilde\rho^2(\mb{r})\label{eq:e_pair_kin}}
does converge.

Various contributions to the total HFB energy as
functions of the cutoff energy are shown in
Fig.~\ref{fig:energy_contribs_fluct.eps}.  The total
energy is stable with respect to $\epsilon_{\text{cut}}$, although some of the
components of the total energy vary significantly. As expected from
Eq.~(\ref{eq:e_pair_kin}), two terms exhibiting large fluctuations
are the kinetic term (with variations of about 2\,MeV) and pairing term
(with variations of about 1.3\,MeV).  Also, the momentum-dependent
spin-orbit term, $E_{S.O.}$, has significant variations of about 1\,MeV. On
the other hand, Skyrme and Coulomb energies are fairly stable with
respect to $\epsilon_{\text{cut}}$.

\section{Numerical Implementation}\label{sec_results}

\subsection{Numerical Framework}

As the pairing renormalization and
regularization procedures
remove the divergent part of the abnormal
density in a different way, one can expect some numerical
differences between both methods. In order to compare their results,
we have performed numerical
calculations using two numerical codes solving the HFB equations:
\begin{itemize}
\item HFBRAD \cite{[Ben05]} --  solves the HFB equations in the spherically symmetric
coordinate basis.  The maximum angular momenta used in calculations
were $j_{\text{max}}=39/2$ for neutrons and $j_{\text{max}}=25/2$ for protons.
\item HFBTHO \cite{[Sto05]} --  diagonalizes the HFB problem in the
axially symmetric transformed harmonic oscillator (HO) basis. Unless
stated otherwise, we use $N_{osc}=$~20 HO shells in the basis.
\end{itemize}

In our calculations, we use the SLy4 \cite{[Cha98]} and SkP \cite{[Dob84]}
parameterizations of the Skyrme functional in the p-h channel and
the contact density-dependent force in the p-p channel, which leads to the pairing energy
density of the form:
\eqa{
%\displaystyle
\mathcal{H}_{pair}({\bf r}) &=& \frac{1}{2}g(\mb{r})\tilde{\rho}(\mb{r})^{2}=\nonumber\\
&=&\frac{1}{2}V_0
\left[1-V_1\left(\frac{\rho(\mb{r})}{\rho_0}\right)~
\right]\tilde{\rho}(\mb{r})^{2} , \label{edhppd}
}
where $\rho_0=0.16$\,fm$^{-3}$. For
$V_1=0$ the resulting pairing interaction is called volume pairing, while
$V_1=1/2$ corresponds to the so-called mixed pairing
prescription (Ref. \cite{[Dob02]} and references quoted therein).

\subsection{Pairing Renormalization}

\begin{figure}
\begin{center}
\includegraphics[width=1.00\columnwidth]{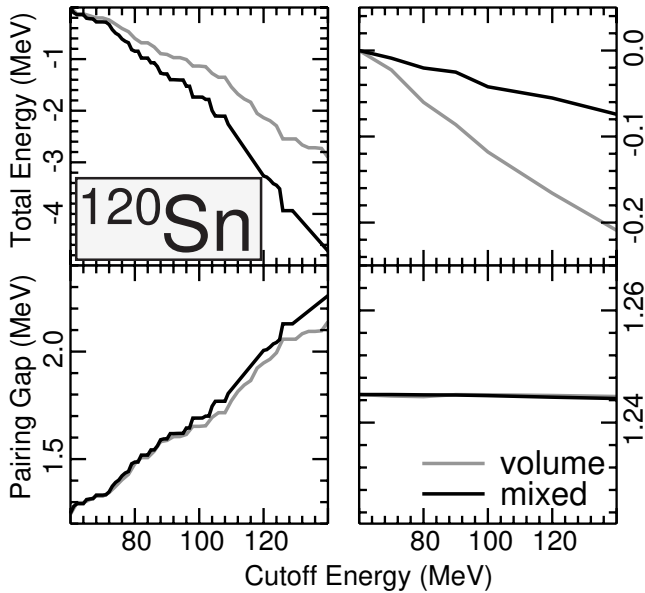}
\end{center}
\caption{Total energy (top) and neutron pairing gap (bottom) in $^{120}$Sn without (left)  and with (right) pairing
renormalization applied.
Results are shown for volume (gray) and mixed (black) pairing.
The total energy
is plotted relative to the values obtained for the cutoff energy
of $\epsilon_{\text{cut}}$=60\,MeV.
}
\label{fig:renorm_no_renorm.eps}
\end{figure}

Figure~\ref{fig:renorm_no_renorm.eps} illustrates the importance of
the pairing renormalization procedure in the case of $^{120}$Sn.
Due to the constraint (\ref{eq:deltabar}) on the pairing strength, the neutron average pairing gap stays by definition constant, while the resulting total energy changes with the cutoff energy
by a few hundred keV. On the other hand,
without pairing renormalization applied, the total energy and the average
neutron gap vary significantly with increasing dimension of the quasiparticle
space. In this case, the total energy changes by several MeV.

\subsection{Pairing Regularization}

\begin{figure}
\begin{center}
\includegraphics[width=1.00\columnwidth]{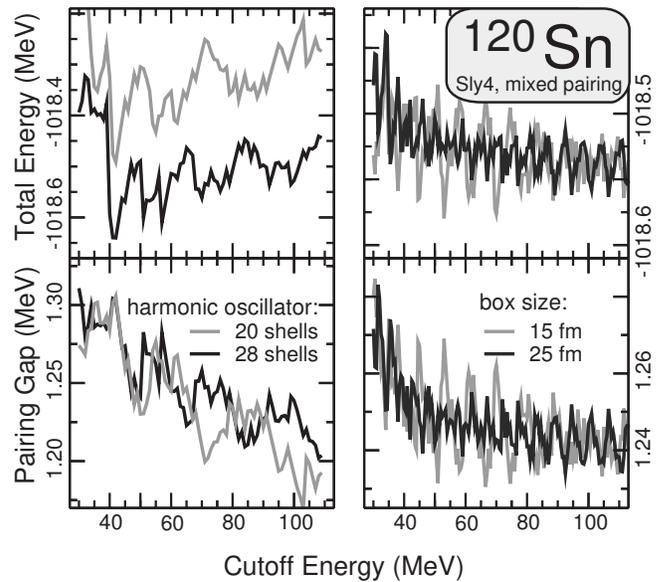}
\end{center}
\caption{Total energy (top) and neutron pairing gap (bottom) in $^{120}$Sn for the two values of $N_{osc}$
(left) or using two box sizes (right). Calculations were performed
using the mixed pairing interaction.}
\label{fig:sn120_2028sh_1525fm.eps}
\end{figure}

The total energy and the average neutron pairing
gap in $^{120}$Sn are shown in
Fig.~\ref{fig:sn120_2028sh_1525fm.eps} after applying the pairing
regularization procedure. The pairing strength $V_0$ is kept constant; it reproduces the neutron pairing gap for $^{120}$Sn at the cutoff energy
of $\epsilon_{\text{cut}}=$60\,MeV.

In the left panels of Fig.~\ref{fig:sn120_2028sh_1525fm.eps}, we show results obtained
in the HO basis, while the results
from the solution of the HFB equations in coordinate space are displayed in the right panels.
One can correlate the coordinate-space
and HO representations by introducing an
`effective box size'  $R\approx \sqrt{2N_{osc}\hbar/m\omega}$
\cite{[Dob96]}. Using this formula, the basis of 20 HO shells
corresponds to a box radius of about 14.5\,fm.
Figure~\ref{fig:sn120_2028sh_1525fm.eps} demonstrates that the
regularization procedure is stable with respect to
the cutoff energy. Moreover, one obtains reasonable results already for
fairly low cutoff energies of about 40\,MeV.
The variations in the total energy in coordinate-space
calculations do not exceed 40\,keV, while they are about 150\,keV
in the HO expansion. The latter number
does not decrease significantly with $N_{osc}$.

The differences in applying the pairing regularization procedure in
the coordinate-space and HO calculations can be explained by the
different way the quasiparticle space is expanded in both
approaches. The particle density $\rho$ is defined by the lower
components of the quasiparticle wave functions, which are localized
within the nuclear interior. On the other hand, the
abnormal density is defined by the products of the upper
and lower components of the quasiparticle wave function. For
the quasiparticle energies that are greater than the modulus of the chemical potential, the upper
components of the quasiparticle wave function are not localized. Therefore, contrary to
the normal density, the abnormal density strongly depends on the
completeness of the s.p. basis outside the nuclear interior.

\begin{figure}
\begin{center}
\includegraphics[width=1.00\columnwidth]{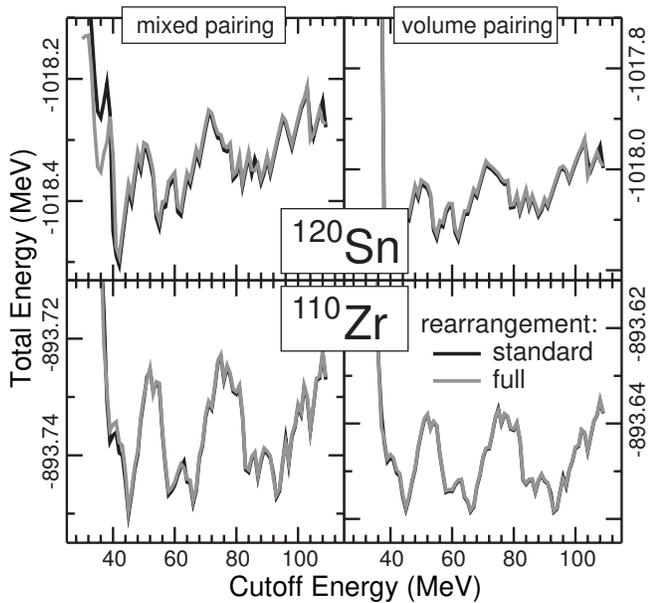}
\end{center}
\caption{Total energy of spherical $^{120}$Sn (top) and deformed $^{110}$Zr
(bottom) obtained with pairing regularization (black
lines) for mixed pairing (left) and volume pairing (right). The results
obtained without the rearrangement terms resulting from the variation
of $k_\text{cut}$ and $k_F$ in Eq.~(\protect\ref{eq:rearr}) are also shown using (gray lines).}
\label{fig:sn120zr110_tho_sly4.eps}
\end{figure}

In the coordinate-space calculations, the box boundary conditions
provide discretization of the spectrum for the quasiparticle
continuum states that are not localized. On the other hand,
all the HO basis states are localized.
Results of stability with
respect to the cutoff energy for the coordinate-space
and HO calculations are, therefore,
different.
As far as the description of nonlocalized states is concerned, the coordinate-space method is superior over the HO expansion method.

Fluctuations in the total energy shown in
Fig.~\ref{fig:sn120_2028sh_1525fm.eps} coincide with $2j+1$-folded degenerate
angular-momentum multiplets of states in spherical nuclei that enter
the pairing window with increasing cutoff energy. This can be
confirmed by performing a similar analysis for a deformed nucleus where
the magnetic degeneracy is lifted. Such results
are shown in Fig.~\ref{fig:sn120zr110_tho_sly4.eps} for deformed $^{110}$Zr
in comparison with spherical $^{120}$Sn.
One can see that the fluctuations of the total energy in $^{110}$Zr are down to about
40\,keV.

The steep increase of the total energy at the cutoff energies below 30\,MeV
results from neglecting quasiparticle states with
significant occupation probability.
This effect is more severe for the
mixed-pairing than for volume-pairing calculations due to the surface-peaked character of mixed pairing fields. On the other hand, the
stability with respect to the cutoff energy is similar in both cases.

We have also tested the importance of the rearrangement terms
arising as a result of the regularization procedure. The gray lines in
Fig.~\ref{fig:sn120zr110_tho_sly4.eps} show results obtained without
taking into account the second and third term of Eq.~(\protect\ref{eq:rearr}).
These terms lead to changes in the total energy
of a few keV and can be safely neglected.

\begin{figure}
\begin{center}
\includegraphics[width=1.0\columnwidth]{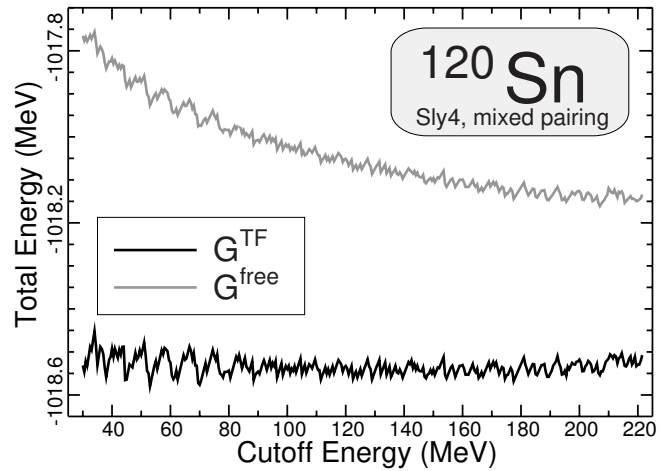}
\end{center}
\caption{Two pairing regularization schemes applied to the case of $^{120}$Sn: the Thomas-Fermi approximation \cite{[Bul02]} (black line)
and the free particle Green's function \cite{[Esb97]} (gray line).
Coordinate-space calculations were performed in a 15\,fm box.}
\label{fig:esbensen_karims_code_2.eps}
\end{figure}

Finally, we have tested the Thomas-Fermi
approximation used in the pairing regularization procedure. Instead
of adopting the Thomas-Fermi ansatz, one can perform regularization
using the free particle Green's function
\cite{[Esb97]}.
As illustrated in Fig.~\ref{fig:esbensen_karims_code_2.eps}, the convergence of the
latter method is very slow; the Thomas-Fermi method is clearly superior.

\subsection{A link between the pairing renormalization and regularization procedures}

\begin{figure}
\begin{center}
\includegraphics[width=0.80\columnwidth]{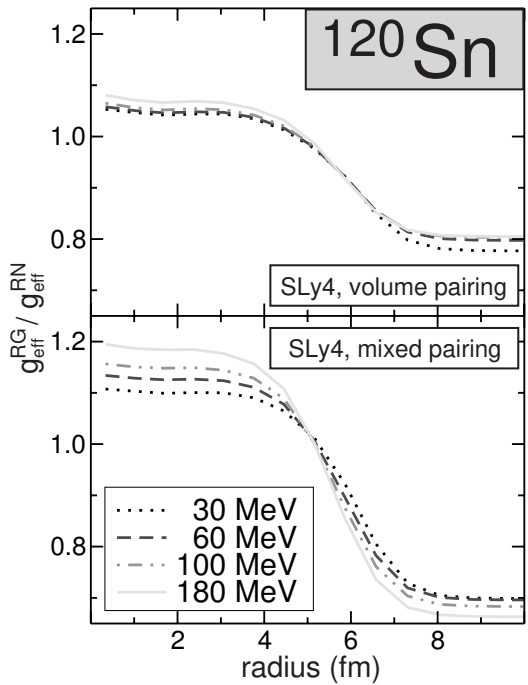}
\end{center}
\caption{Ratio between the effective pairing strengths for pairing
regularization and renormalization, $g_{eff}^{RG}/g_{eff}^{RN}$, for
the volume (upper panel) and mixed pairing (lower panel) in
$^{120}$Sn for several values of $\epsilon_{\text{cut}}$.} \label{fig:vk_v0.eps}
\end{figure}

The renormalized
and regularized pairing calculations are based, in fact, on
two different effective interactions. Consequently,
their results should be comparable only as much as their effective
pairing strengths $g_{eff}$ are similar.
By expanding Eq. (\ref{eq:g^ri}) at very high
cutoff energies  ($k_F/k_\text{cut}<<1$), one obtains:
\eq{g_{eff}(\mb{r})\approx\left(1-\frac{M^*(\mb{r})g(\mb{r})}{2\pi^2\hbar^2}
k_\text{cut}(\mb{r})\right)^{-1}g(\mb{r}), }
which has the
form of $g_{eff}=\alpha g$. For the volume
pairing, the proportionality factor $\alpha$ is $\rho$-dependent only through the weak
density dependence of the effective mass $M^*$. On the other hand, for the mixed pairing,
it also depends on $\rho$ through the density dependence of $g$.
Therefore, while for the volume pairing
the renormalization procedure may be considered as a fair approximation to the
regularization scheme, this is not the case for
the mixed pairing,
or -- more generally -- for any density-dependent pairing. Still, this approximate equality
of the effective pairing strengths for the pairing regularization and renormalization is
an explanation
of the remarkable stability of the total energy in phenomenological pairing renormalization
treatment (see Fig.~\ref{fig:renorm_no_renorm.eps}), and it also explains why results obtained
for the volume pairing are more stable than those in the mixed pairing variant.

This effect can be clearly seen in Fig.~\ref{fig:vk_v0.eps}. The
ratio between the effective pairing strengths in the regularization and
renormalization methods is much closer to unity for the
volume pairing than for the mixed pairing in the region of space, where the pairing energy density is maximal.

\section{Comparison between pairing renormalization and
regularization procedures}\label{chains}

In this section, we present a comparison between pairing
renormalization and regularization procedures applied
to a large number of nuclei. As representative
results, we discuss results obtained for the drip-to-drip line
isotopic chains of spherical Sn nuclei as well as for deformed Dy
nuclei. Calculations are performed for the
volume and/or mixed pairing interactions by using the HFB+THO approach.

\subsection{Spherical Nuclei}

\begin{figure*}[htb]
\includegraphics[width=1.9\columnwidth]{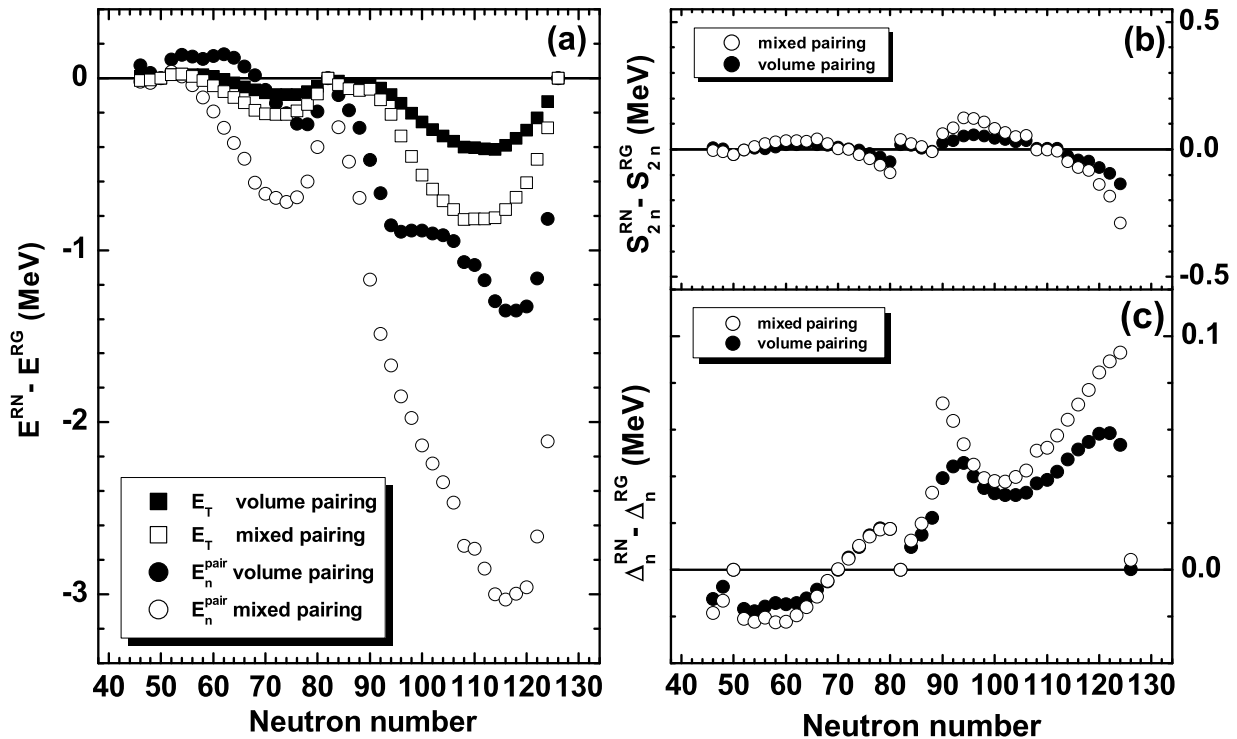}
\caption{Differences between pairing renormalization (RN) and
regularization (RG) procedures for (a) total
energies and neutron pairing energies, (b) two neutron separation
energies, and (c) the average neutron gaps. The HFB+THO calculations are
performed for the chain of the spherical Sn isotopes using the SkP
Skyrme parameterization.} \label{fig1m}
\end{figure*}

Figure \ref{fig1m} displays differences between the pairing renormalization and regularization
procedures for the Sn isotopes. Calculations are performed with both volume
and mixed pairing interactions.
For the two-neutron separation energies, the
maximum difference between the renormalization and regularization schemes is about 100 (300)\,keV for the volume (mixed)
pairing. In the neutron gap, the corresponding difference is
about 50 (100)\,keV, and in nuclear radii (not displayed) it is practically
negligible (about 0.01\,fm). The largest differences show up in pairing energies -- about 1 (3) MeV for the volume (mixed)
pairing; however, total energy differences are much smaller -- about 400
(800)\,keV.

\begin{figure*}[htb]
\includegraphics[width=1.60\columnwidth]{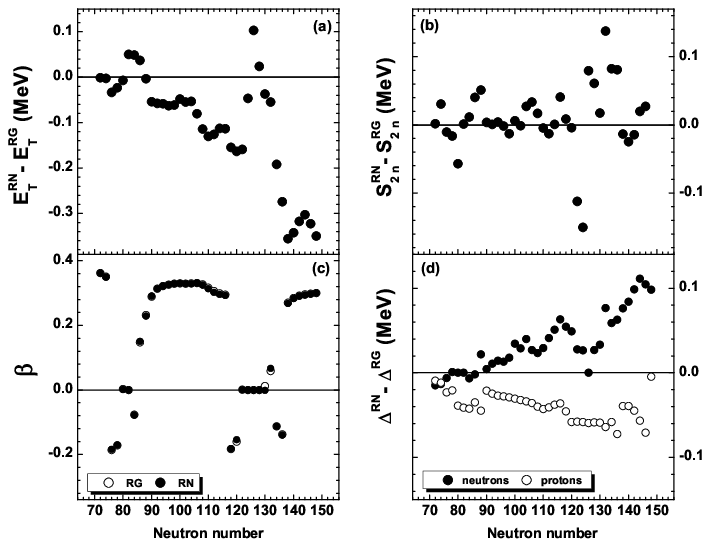}
\caption{Similar to Fig.~\protect\ref{fig1m} except for the deformed Dy isotopes. Quadrupole deformations are displayed in panel (c). Mixed pairing interaction was used.} \label{fig2m}
\end{figure*}

Analyzing the total energies obtained in both methods,
Fig.~\ref{fig1m}(a), one can see that the pairing renormalization
procedure gives systematically more binding. The differences are
negligible for stable nuclei and nuclei near the proton
drip line. They increase in mid-shell nuclei near the two-neutron
drip line where the pairing effects are the largest, and then
decrease towards the closed-shell nucleus $^{176}$Sn located just at
the two-neutron drip line. In general, both procedures give more similar results
in the case of volume pairing than in the case
of mixed pairing.

Recently, the pairing regularization procedure has been analyzed in the
context of relativistic mean-field approximation \cite{[Nik05]}.
In order to simulate the finite
range contribution to the nuclear matter pairing gap coming from the
Gogny pairing force,  it
was necessary to introduce strong density dependence in the pairing
strength of the contact interaction.

Using the regularization procedure and calculating the Sn chain with
both volume and newly constructed (surface) contact interaction, the
authors of Ref.~\cite{[Nik05]} have found differences in pairing
energies of the order of 20\,MeV in the neutron-rich nuclei around
$^{148}$Sn. In our work, for the same nuclei, the differences in pairing energies
between volume and mixed pairing variants do not exceed 2.6\,MeV. This comparison shows that
the density-dependent contact interaction proposed in Ref.~\cite{[Nik05]} is
questionable for finite nuclei, despite its agreement with
the finite-range Gogny pairing force in the infinite nuclear matter.

\subsection{Deformed Nuclei}

We applied the pairing renormalization and regularization
procedures for the chain of deformed Dy isotopes. Differences between both sets of
results are shown in
Fig.~\ref{fig2m}.
We show only the results with the mixed pairing, since, as in the spherical nuclei, the
differences between both procedures are larger in this case.

As seen in Fig.~\ref{fig2m}(c), most of the nuclei considered are well deformed, and the deformations are practically the same
within both procedures. Despite the fact that the maximum difference
in the pairing energy is around 3\,MeV (not shown), other quantities are very similar. The maximum difference in the total energy is
about 360\,keV, in the two-neutron separation energy -- 160\,keV, in
the pairing gaps -- 110\,keV, and in the rms radii (not shown) -- less than 0.005\,fm.

\section{Summary and Conclusions}\label{sec_summary}

In this work, we investigated the pairing regularization method using
the s.p.\ Green's function in the Thomas-Fermi approximation and
found it to be very suitable for a description of spherical and
deformed nuclei. We checked the stability of the method with respect to the
cutoff energy and found fluctuations in the total energy
below 200\,keV. Fluctuations coming from the method itself
do not exceed 50\,keV for the cutoff energy as low as 30\,MeV. However, if
a still lower cutoff energy is assumed, the Thomas-Fermi approximation to the s.p. Green's function may no longer be
valid.

We found that
the differences between pairing renormalization and
regularization procedures for volume and mixed pairing are rather small. Therefore, we conclude
that physical conclusions previously obtained within the pairing renormalization
scheme remain valid.  Nevertheless, we believe that the theoretical
motivation and simplicity of the regularization method is preferred
to a phenomenological renormalization scheme.

\begin{acknowledgments}
Discussions with Aurel Bulgac are gratefully acknowledged. This work
was supported in part by the U.S.\ Department of Energy under
Contract Nos.\ DE-FG02-96ER40963 (University of Tennessee),
DE-AC05-00OR22725 with UT-Battelle, LLC (Oak Ridge National
Laboratory), by the National Nuclear Security Administration under
the Stewardship Science Academic Alliances program through DOE
Research Grant DE-FG03-03NA00083;  by the Polish Committee for
Scientific Research (KBN) under contract N0.~1~P03B~059~27 and  by
the Foundation for Polish Science (FNP).
\end{acknowledgments}

\appendix{}

\section{Divergence in the Abnormal Density}\label{app:divergence}

In the DFT-HFB approach, the starting point is the Energy Density
Functional (EDF) $\mathcal{H}[\rho,\tilde\rho]$, where $\rho$ is the
particle density and $\tilde\rho$ is the abnormal density:
\eqa{
\rho(\mb{r_2}\sigma_2\tau_2,\mb{r_1}\sigma_1\tau_1)&\!\!=\!\!&
\langle\Phi|a^{\dagger}_{\mb{r_1}\sigma_1\tau_1}a_{\mb{r_2}\sigma_2\tau_2}|\Phi\rangle,\\
\tilde\rho(\mb{r_2}\sigma_2\tau_2,\mb{r_1}\sigma_1\tau_1)&\!\!=\!\!&
-2\sigma_1\langle\Phi|a_{\mb{r_1}-\sigma_1\tau_1}a_{\mb{r_2}\sigma_2\tau_2}|\Phi\rangle,
}
where $a$ and $a^{\dagger}$ are the particle annihilation and
creation operators, respectively, and $|\Phi\rangle$ is the HFB
state. In the following, we assume that $|\Phi\rangle$ is a product of
the neutron and proton states, $|\Phi_{\nu}\rangle|\Phi_{\pi}\rangle$.
Therefore, the neutron and proton wave functions are not coupled, and
in the notation below we can, for simplicity, omit the isospin index
with the understanding that all equations are separately valid for
neutrons and protons.

For the HFB state $|\Phi\rangle$, the particle and abnormal
densities can be written as \cite{[Dob84]}:
\eqa{
\rho(\mb{r_2}\sigma_2,\mb{r_1}\sigma_1)&=&\sum_{E_i>0}\varphi_{2i}(\mb{r_2}\sigma_2)\varphi_{2i}^*(\mb{r_1}\sigma_1),\\
\tilde\rho(\mb{r_2}\sigma_2,\mb{r_1}\sigma_1)&=&-\sum_{E_i>0}\varphi_{2i}(\mb{r_2}\sigma_2)\varphi_{1i}^*(\mb{r_1}\sigma_1),\label{abnormal}
}
where the two-component quasiparticle wave function
$\varphi$ is the solution of the HFB equation:
\eqa{
&&\sum_{\sigma_1}\int d^3\mb{r_1}\left[\begin{array}{cc}h_{\mu}(\mb{r_2}\sigma_2,\mb{r_1}\sigma_1)&\tilde h(\mb{r_2}\sigma_2,\mb{r_1}\sigma_1)\\\tilde h(\mb{r_2}\sigma_2,\mb{r_1}\sigma_1)&-h_{\mu}(\mb{r_2}\sigma_2,\mb{r_1}\sigma_1)\end{array}\right]\times\nonumber\\
&&\times\left[\begin{array}{c}\varphi_{1i}(\mb{r_1}\sigma_1)\\\varphi_{2i}(\mb{r_1}\sigma_1)\end{array}\right]=E_i\left[\begin{array}{c}\varphi_{1i}(\mb{r_2}\sigma_2)\\\varphi_{2i}(\mb{r_2}\sigma_2)\end{array}\right],\label{app:HFB_eq}}
for a given quasiparticle energy $E_i$.

The HFB equations are a result of variational minimization of the
energy density functional $\mathcal{H}[\rho,\tilde\rho]$ with the constraint of the
mean value of particles kept constant:
\eq{\left.\delta\mathcal{H}\right|_{\langle \hat N\rangle=N}=0.}
This condition defines the s.p.\ Hamiltonian $h_{\mu}$ and
the pairing Hamiltonian $\tilde h$ in the HFB equations (\ref{app:HFB_eq}):
\eqa{
\lefteqn{h_{\mu}(\mb{r_2}\sigma_2,\mb{r_1}\sigma_1)=\frac{\delta\mathcal{H}[\rho,\tilde\rho]}{\delta\rho(\mb{r_1}\sigma_1,\mb{r_2}\sigma_2)}-\mu}\\
&=&-{\mb{\nabla}_{\mb{r_2}}}\frac{\hbar^2}{2M^*(\mb{r_2}\sigma_2,\mb{r_1}\sigma_1)}{\mb{\nabla}_{\mb{r_1}}}
+U(\mb{r_2}\sigma_2,\mb{r_1}\sigma_1)-\mu\label{app:hmu_def}\nonumber \\
&&\lefteqn{\tilde h(\mb{r_2}\sigma_2,\mb{r_1}\sigma_1)=\frac{\delta\mathcal{H}[\rho,\tilde\rho]}{\delta\tilde\rho(\mb{r_1}\sigma_2,\mb{r_2}\sigma_2)}}\label{app:delta_def},}
where $M^*$ is the effective mass and $U$ is the self-consistent mean-field potential.
In the following derivations, the spin-orbit term is omitted as unimportant in the regularization scheme, although it is, of course, always included in calculations.

By multiplying the HFB equations (\ref{app:HFB_eq}) by vector
$[\varphi_{2i}^*,-\varphi_{1i}^*]$, integrating over coordinates and
summing over all the positive energy HFB solutions, one obtains:
\begin{widetext}
\eqa{
\lefteqn{\sum_{E_i>0,\sigma_2} E_i\int d^3\mb{r_2}
\left[\begin{array}{cc}\varphi_{2i}^*(\mb{r_2}\sigma_2),&-\varphi_{1i}^*(\mb{r_2}\sigma_2)
\end{array}\right]
\left[\begin{array}{c}\varphi_{1i}(\mb{r_2}\sigma_2)\\\varphi_{2i}(\mb{r_2}\sigma_2)
\end{array}\right]=}&&\nonumber\\
&&=\sum_{E_i>0,\sigma_1\sigma_2} \iint d^3\mb{r_1}d^3\mb{r_2}
\left[\begin{array}{cc}\varphi_{2i}^*(\mb{r_2}\sigma_2),&-\varphi_{1i}^*(\mb{r_2}\sigma_2)
\end{array}\right]
\left[\begin{array}{cc}h_{\mu}(\mb{r_2}\sigma_2,\mb{r_1}\sigma_1)&\tilde h(\mb{r_2}\sigma_2,\mb{r_1}\sigma_1)\\
\tilde h(\mb{r_2}\sigma_2,\mb{r_1}\sigma_1)&-h_{\mu}(\mb{r_2}\sigma_2,\mb{r_1}\sigma_1)
\end{array}\right]
\left[\begin{array}{c}\varphi_{1i}(\mb{r_1}\sigma_1)\\\varphi_{2i}(\mb{r_1}\sigma_1)
\end{array}\right]
}
i.e.,
\eqa{
\lefteqn{\sum_{E_i>0,\sigma_1} E_i \int d^3\mb{r_1}
\left\{\varphi_{2i}^*(\mb{r_1}\sigma_1)\varphi_{1i}(\mb{r_1}\sigma_1)
-\varphi_{1i}^*(\mb{r_1}\sigma_1)\varphi_{2i}(\mb{r_1}\sigma_1)\right\}=}\nonumber\\
&&=\sum_{E_i>0,\sigma_1\sigma_2} \iint d^3\mb{r_1}d^3\mb{r_2}
\bigg\{\varphi_{2i}^*(\mb{r_2}\sigma_2)h_{\mu}(\mb{r_2}\sigma_2,\mb{r_1}\sigma_1)\varphi_{1i}(\mb{r_1}\sigma_1)
+\varphi_{2i}^*(\mb{r_2}\sigma_2)\tilde h(\mb{r_2}\sigma_2,\mb{r_1}\sigma_1)\varphi_{2i}(\mb{r_1}\sigma_1)+\nonumber\\
&&+\varphi_{1i}^*(\mb{r_2}\sigma_2)h_{\mu}(\mb{r_2}\sigma_2,\mb{r_1}\sigma_1)\varphi_{2i}(\mb{r_1}\sigma_1)
-\varphi_{1i}^*(\mb{r_2}\sigma_2)\tilde h(\mb{r_2}\sigma_2,\mb{r_1}\sigma_1)\varphi_{1i}(\mb{r_1}\sigma_1)\bigg\}.\label{app:eq_big}
}
Since for every HFB solution $([\varphi_{1i},\varphi_{2i}],E_i)$ there
exists also an orthogonal solution $([\varphi_{2i},-\varphi_{1i}],-E_i)$,
the left-hand side of Eq.~(\ref{app:eq_big}) vanishes as a sum over
scalar products of orthogonal wave functions.

For local and spin-independent Hamiltonians $h_{\mu}$ and $\tilde h$,
Eqs.~(\ref{app:hmu_def}) and (\ref{app:delta_def}) read
\eqa{
h_{\mu}(\mb{r_2}\sigma_2,\mb{r_1}\sigma_1)&=&
-\mb{\nabla}_{\mb{r_2}}\frac{\hbar^2}{2M^*(\mb{r_2})}\delta(\mb{r_2}-\mb{r_1})
\delta_{\sigma_2,\sigma_1}\mb{\nabla}_{\mb{r_1}}
+(U(\mb{r_2})-\mu)\delta(\mb{r_2}-\mb{r_1})\delta_{\sigma_2,\sigma_1},\label{app:loc2}\\
\tilde h(\mb{r_2}\sigma_2,\mb{r_1}\sigma_1)&=&
\tilde h(\mb{r_2})\delta(\mb{r_2}-\mb{r_1})\delta_{\sigma_2,\sigma_1}.\label{app:loc1}
}
Note that for an attractive pairing force, the local pairing potential
$\tilde h(\mb{r})=-\Delta(\mb{r})$ is negative,
where $\Delta(\mb{r})$ is the standard position-dependent pairing gap.
By defining function $\mathcal F_{\epsilon_{\text{cut}}}$ as
\eq{\sum_{E_i>0,\sigma} \left[\varphi_{1i}(\mb{r_2}\sigma)\varphi_{1i}^*(\mb{r_1}\sigma)
+\varphi_{2i}(\mb{r_2}\sigma)\varphi_{2i}^*(\mb{r_1}\sigma)\right]
=\mathcal F_{\epsilon_{\text{cut}}}(\mb{r_2}-\mb{r_1}).}
and using expression (\ref{abnormal}) for the abnormal density, one obtains
after integrating the kinetic-energy term by parts:
\eqa{0&=&-\int d^3\mb{r_1}d^3\mb{r_2}\delta(\mb{r_2}-\mb{r_1})
\left[\tilde h(\mb{r_2})\left[\mathcal F_{\epsilon_{\text{cut}}}(\mb{r_2}-\mb{r_1})
-2\rho(\mb{r_2},\mb{r_1})\right]
+\left(\frac{\hbar^2}{2M^*}\mb{\nabla}_{\mb{r_2}}\mb{\nabla}_{\mb{r_1}}
+U(\mb{r_2})-\mu\right)2\tilde\rho(\mb{r_1},\mb{r_2})\right]=\nonumber\\
&=&-\int d^3\mb{r}d^3\mb{x}\delta(\mb{x})\left[\tilde h(\mb{r})
\left[\mathcal F_{\epsilon_{\text{cut}}}(\mb{x})
-2\rho(\mb{r},\mb{r})\right]
 +\left(\frac{\hbar^2}{2M^*}\left(\frac{1}{4}\nabla_{\mb{r}}^2-\nabla_{\mb{x}}^2\right)
+U(\mb{r})-\mu\right)2\tilde\rho(\mb{r}-\mb{x}/2,\mb{r}+\mb{x}/2)\right],\label{app:eq_big2}}
\end{widetext}
where
\eqa{
\mb{r}&=&\frac{\mb{r_1}+\mb{r_2}}{2},\\
\mb{x}&=&\mb{r_2}-\mb{r_1},}
and
\eqa{
\rho(\mb{r_2},\mb{r_1})&=&\sum_{\sigma}\rho(\mb{r_2}\sigma,\mb{r_1}\sigma),\\
\tilde\rho(\mb{r_2},\mb{r_1})&=&\sum_{\sigma}\tilde\rho(\mb{r_2}\sigma,\mb{r_1}\sigma).
}
When the summation over positive quasiparticle energies is extended to infinity,
the completeness relation implies that
\eq{\mathcal F_{\epsilon_{\text{cut}}}(\mb{r_2}-\mb{r_1})=\delta(\mb{r_2}-\mb{r_1}),}
and the only term in Eq.~(\ref{app:eq_big2}) capable of canceling out this singularity
is $\nabla_{\mb{x}}^2\tilde\rho(\mb{r}-\mb{x}/2,\mb{r}+\mb{x}/2)$. Therefore,
the Laplacian of the abnormal density
$\nabla_{\mb{x}}^2\tilde\rho(\mb{r}-\mb{x}/2,\mb{r}+\mb{x}/2)$ must be singular
at $\mb{x}=0$. Moreover, using the expression
\eq{\nabla^2\frac{1}{|\mb{r}|}=-4\pi\delta(\mb{r}),}
it is clear that due to the zero-range pairing interaction
abnormal density $\tilde\rho$ has an ultraviolet $1/x$ divergence:
\eq{\left.\tilde\rho(\mb{r}-\mb{x}/2,\mb{r}+\mb{x}/2)\sim-\frac{\tilde h(\mb{r})
M^*(\mb{r})}{4\pi\hbar^2|\mb{x}|}\right|_{\mb{x}\rightarrow 0}.}

%\bibliographystyle{bib_pjpb}
%\bibliography{pjpb}

\end{document}